\documentclass{IEEEtran}
\usepackage{amsmath,amssymb,enumitem,mathtools}
\usepackage{algpseudocode,algorithm}
\usepackage{xcolor}
\usepackage{subfigure}
\usepackage{microtype}
\usepackage{booktabs}
\usepackage{cite}
\usepackage{graphicx}
\graphicspath{{./figures/}}

\usepackage{float}

\usepackage{epstopdf}

\algrenewcommand\algorithmicindent{2em}

\algblockdefx{MRepeat}{EndRepeat}{\textbf{repeat}}{}
\algnotext{EndRepeat}

\newcounter{proxy}[table]
\newcommand{\subjto}{\text{subject to}}
\newcommand{\prox}{\text{prox}}
\DeclareMathOperator*{\mmz}{\text{minimize}}
\DeclareMathOperator*{\argmin}{arg\,min}

\title{Fast non-coplanar beam orientation optimization based on group sparsity}

\author{Daniel~O'Connor, Yevgen~Voronenko, Dan~Nguyen, Wotao~Yin, and~Ke~Sheng}

\date{}

\begin{document}
\maketitle
\begin{abstract}

\,\emph{Objective:} The selection of beam orientations, which is a key step in radiation treatment planning,
 is particularly challenging for non-coplanar radiotherapy systems
due to the large number of candidate beams.
In this paper, we report progress on the group sparsity approach to beam orientation optimization,
wherein beam angles are selected by solving a large scale fluence map optimization problem
with an additional group sparsity penalty term that encourages most candidate beams to be inactive.
\emph{Methods:} The optimization problem is solved using an accelerated
proximal gradient method, the Fast Iterative Shrinkage-Thresholding Algorithm (FISTA).
We derive a closed-form expression for a relevant proximal operator which enables the application
of FISTA.
The proposed algorithm is used to create non-coplanar
treatment plans for four cases (including head and neck,
lung, and prostate cases), and the resulting plans are
compared with clinical plans.
\emph{Results:} The dosimetric quality of the group sparsity treatment plans 
is superior to that of the clinical plans. 
Moreover, the runtime for the group sparsity approach is typically about 5 minutes.
Problems of this size could not be handled using the previous group sparsity method
for beam orientation optimization, which was slow to solve much smaller coplanar cases.
\emph{Conclusion/Significance:} This work demonstrates for the first time that the group sparsity approach, 
when combined with an accelerated proximal gradient method
such as FISTA, works effectively for 
\emph{non-coplanar} cases with 500-800 candidate
beams.
  
\end{abstract}


\maketitle 

\section{Introduction}

\PARstart{I}{n} current radiation therapy planning practice, the beam orientation
is most commonly set up manually before fluence map optimization.  For coplanar
planning, the need for such a step was partially alleviated by emerging
arc therapy.  However, the challenge still exists for non-coplanar
radiotherapy where manual beam selection is unintuitive and
impractical. It has been shown that manually selected non-coplanar arcs are not
superior to coplanar arc therapy for liver SBRT and are substantially
inferior to beam orientation optimized static beam non-coplanar plans
\cite{woods2016viability}.

Due to the size and the
combinatorial nature of the problem, beam orientation optimization algorithms usually alternate between
beam angle selection and fluence map optimization steps: at each stage,
a fluence map optimization problem is solved (using the beam angles that have been selected so far),
and then a new beam is added to the collection according to various heuristics
\cite{romeijn2005column, breedveld2012icycle,lim2008iterative,bertsimas2013hybrid,yarmand2013two,rocha2015does,bangert2016accelerated}.
A notable example is the algorithm based on column generation \cite{romeijn2005column,dong20134pi,dong2014feasibility}.
While these methods have proved to be useful, their runtimes do not scale well with the number of beams to be selected,
because ever larger fluence map optimization subproblems must be solved at successive iterations as more beams are added to the collection.
For example, \cite{bangert2016accelerated} notes that the time required to select $9$ beams is substantially longer than the
time to select $7$ beams.
This is particularly a concern for non-coplanar IMRT, where it has been found that more beams can be utilized
during treatment before hitting a point of dosimetric diminishing returns \cite{dong20134pi_2}.
Moreover, these interleaved algorithms select beam angles one by one in a greedy manner, and we might hope for a more holistic approach in which
all beam angles are selected at once.

An elegant alternative approach, based on \emph{group sparsity}, was presented in \cite{jia2011beam}.
If computational resources were unlimited, it would be natural to select beam angles by
solving a fluence map optimization problem involving a large number of candidate beams, with 
a constraint on the number of beams which are allowed to be active.
Equivalently, the constraint on the number of active beams could be replaced by a penalty
term in the objective function which is proportional to the number of active (nonzero) beams.
Of course, the resulting optimization problem is non-convex and computationally intractable.
In the group sparsity approach, the $\ell_{2,1}$-norm of the fluence map $x$, defined by $\| x \|_{2,1} = \sum_b \| x_b \|_2$
(where $x_b$ is the fluence map for beam $b$),
is used as a \emph{convex surrogate} for the non-convex beam counting function,
just as the $\ell_1$-norm is used as a surrogate for the $\ell_0$-penalty to promote sparsity
in compressed sensing problems.
The group sparsity penalty term encourages most candidate beams to be inactive,
and the remaining active beams are the ones selected to be used during treatment.  
This method has at least a theoretical
appeal, in that beam angles are selected not in a greedy manner but instead by finding the global
minimizer for a convex optimization problem.  The method connects to a large literature on sparsity
and group sparsity as it is used in areas such as signal processing, statistics, compressed sensing, and machine learning
\cite{bach2012optimization,simon2013sparse,meier2008group,bach2008consistency,huang2010benefit}. 
Unfortunately, the method as presented
in \cite{jia2011beam} was rather slow, with runtimes of several hours reported for \emph{coplanar}
head and neck cases
involving just $72$ candidate beams.  
The paper concluded that more work was necessary to make the group sparsity approach
tractable for non-coplanar beam angle selection, where we are faced with $800$ or so candidate beams.

In this paper we report progress on the group sparsity line of inquiry for beam orientation optimization.
The group sparsity penalized problem is expressed in a form that is suitable for an accelerated
proximal gradient method, the Fast Iterative Shrinkage-Thresholding Algorithm (FISTA) \cite{beck2009FISTA},
and an efficient closed-form expression is derived for the required proximal operator.
The $O(1/k^2)$ convergence rate of FISTA is a dramatic
improvement on the $O(1/k)$ convergence rate of the standard proximal gradient method,
also known as forward-backward method, which was used in \cite{jia2011beam}.
(Here $k$ is the iteration number. This difference in convergence rates is illustrated in figure~\ref{convergencePlot}
in section~\ref{numericalResults}.)
The effectiveness of the resulting algorithm for \emph{non-coplanar} beam orientation optimization
is demonstrated by using the algorithm to create non-coplanar treatment plans for four patients,
including head and neck, lung, and prostate patients. 
A preliminary version of this work has been reported in the conference abstract~\cite{aapm2016talk}.

\section{Methods}
\subsection{Problem formulation} 
Starting from a large collection of $B$ candidate beams, we select beams to be used during treatment
by minimizing the convex objective function
\begin{align}
\label{generalFmoProb}
\notag \underbrace{\frac12\|(\ell - A_0 x)_+\|_2^2}_{\text{PTV}}  
&+ \underbrace{\sum_{i=0}^N \frac{\alpha_i}{2} \| (A_i x - d_i)_+\|_2^2 + \frac{\beta_i}{2} \| A_i x \|_2^2}_{\text{OARs}} \\
&+ \underbrace{\gamma \| Dx \|_1^{(\mu)}}_{\text{smoothness}} + \underbrace{\sum_{b=1}^B w_b \| x_b \|_2}_{\text{group sparsity}} 
\end{align}
subject to the constraint that $x \geq 0$, 
where:
\begin{itemize}
\item
$x_b$ is the vector of beamlet intensities for beam $b$.
\item The optimization variable $x$ is the concatenation of the vectors $x_b$.
\item $N$ is the number of organs at risk (OARs).
\item The notation $y_+$ denotes $\max(y,0)$, with the maximum interpreted componentwise.
\item The matrices $A_i$ are the dose-calculation matrices
for the planning target volume ($i = 0$) and for the organs at risk ($i = 1,\ldots,N$).
\item
The matrix $D$ represents a discrete gradient operator, so that $Dx$ is a list of intensity
differences between adjacent beamlets.
\item
The function $\| \cdot \|_1^{(\mu)}$ is the Huber penalty (with parameter $\mu > 0$), defined by
\begin{align}
\label{huberDef}
\| y \|_2^{(\mu)} &= \sum_j | y_j |^{(\mu)}, \\
\notag \qquad | y_j |^{(\mu)} &= 
\begin{cases} 
\frac{1}{2\mu} y_j^2 & \quad \text{if } |y_j| \leq \mu, \\
|y_j| - \frac{\mu}{2} & \quad \text{otherwise}.
\end{cases}
\end{align}
(The notation $\| \cdot \|_1^{(\mu)}$ reminds us that the Huber penalty is a smoothed out version
of the $\ell_1$-norm, and $\mu$ controls the amount of smoothing.)
\end{itemize}
Without the group sparsity term, problem~\eqref{generalFmoProb}
would be a standard fluence map optimization problem.
The term $(1/2) \| (\ell - A_0 x)_+ \|_2^2$ encourages a prescribed minimum
dose of radiation (stored in the vector $\ell$) to be delivered to the PTV, 
while the terms $(\alpha_i/2) \|(A_i x - d_i)_+\|_2^2$
encourage the radiation delivered to the PTV and OARs not to exceed
prescribed maximum doses (stored in the vectors $d_i$).  
The terms $(\beta_i/2) \| A_i x \|_2^2$ provide additional control over
dose delivered to the OARS.  (We always set $\beta_0 = 0$.)
The regularization term $\gamma \| Dx \|_1^{(\mu)}$, which is a smoothed
total variation penalty, encourages piecewise-smooth fluence maps.
More details about the beam setup and dose calculation are provided
in section~\ref{setupSection}.

The group sparsity approach to beam angle selection is based on the following
fact: \emph{the $\ell_{2,1}$-norm penalty encourages most candidate beams to have $x_b$ identically zero.}
Upon solving problem~\eqref{generalFmoProb}, we find that only a small number of candidate beams
are nonzero, and these are the beams selected to be used during treatment.
The $\ell_{2,1}$-norm can be viewed as a convex surrogate for the $\ell_{2,0}$-penalty
which counts the number of nonzero groups (beams) in $x$, just as the $\ell_1$-norm is
ubiquitous as a convex surrogate for the $\ell_0$-penalty
which computes the number of nonzero components of $x$.
Promoting group sparsity by penalizing the $\ell_{2,1}$-norm is a popular technique
in areas such as statistics, machine learning, and signal processing
\cite{bach2012optimization,simon2013sparse,meier2008group,bach2008consistency,huang2010benefit}.

\paragraph{Selecting the weights $w_b$ in the group sparsity term}
Some beams must only travel a short distance through the body to reach the PTV,
whereas other ``long path'' beams must travel a greater distance through the body
before reaching the PTV.  
To overcome attenuation, a ``long path'' beam must be fired more intensely
than a short path beam in order to deliver the same dose to the PTV.
If all the weights $w_b$ in the group sparsity term are chosen to be equal, 
then the group sparsity penalty
introduces a bias in favor of short path beams, because a long path
beam $b$ requires $\| x_b \|_2$ to be large in order to target the PTV effectively.
We choose the weights $w_b$ to compensate for this bias.

Let $n_b$ be the number of beamlets in beam $b$ with a trajectory that intersects the PTV.
Suppose that beam $b$ is fired uniformly, so that $x_b = \lambda \vec 1$,
and the scalar $\lambda$ is chosen so that the mean dose delivered to the PTV
by beam $b$ is $1$ Gy.  Then it is easy to check that 
$\| x_b \|_2 = \sqrt{n_b}/\text{mean}(A_0^b \vec 1 \,)$,
where $A_0^b$ is the dose-calculation matrix from beam $b$
to the PTV.  We choose the weights $w_b$
so that
\begin{equation}
\label{weightFormula}
w_b = \frac{c \cdot \text{mean}(A_0^b \vec 1 \,)}{\sqrt{n_b}}.
\end{equation}
The scalar $c$ is chosen to be the same for all beams,
and $c$ is tuned by trial and error to achieve the desired
group sparsity level.

\subsection{Optimization algorithm}
In this section we assume familiarity with the definition of the proximal operator (also referred to as prox-operator),
as well as the proximal gradient method from convex optimization and an accelerated version of the proximal
gradient method known as FISTA.  These topics are reviewed in appendix~\ref{fistaSection}. 
An accessible introduction to proximal algorithms
can be found in \cite{parikh2013proximal}; see also \cite{chambolle2016introduction,236cNotes,combettes2011proximal}.
\label{booFistaSection}
\paragraph{Preliminary remarks}
Problem \eqref{generalFmoProb} is difficult to solve because the group sparsity penalty
is \emph{nondifferentiable}.  
Importantly, this rules out the direct use of quasi-Newton methods, which require the objective function to be differentiable.
An additional difficulty is that the dose-calculation matrices $A_i$ are very large,
due to the large number of candidate beams.  (There may be 500-800 candidate beams for a non-coplanar
beam angle selection problem, whereas standard non-coplanar fluence map optimization problems involve only 10-20 beams.)
Despite the fact that the matrices $A_i$ are sparse, they take up many gigabytes of
computer memory.  For example, in case ``LNG\#1'' discussed below (see section~\ref{setupSection}), the matrix $A$ obtained
by stacking the matrices $A_i$ has dimensions
$57258 \times 90656$.  Only $5.75\%$ of the entries of $A$ are nonzero,
but still $A$ takes up about $9.5$ gigabytes of computer memory.
(And this is after downsampling.)  This prevents us from solving problem \eqref{generalFmoProb}
using classical interior point methods, which typically have quadratic memory complexities
and cubic arithmetic complexities.  At each iteration, an interior point method would require
solving a linear system of equations involving the matrices $A_i$, and solving
such a large linear system is computationally intractable.  In recent years, much
research in convex optimization has focused on
a class of algorithms known as ``proximal algorithms'', which are well suited
to this type of large scale, nondifferentiable, constrained convex optimization problem.
But, even within the class of proximal algorithms, it is difficult to find a method which is
capable of solving problem \eqref{generalFmoProb} efficiently.
One of the most popular proximal algorithms, the alternating direction method of multipliers (ADMM) \cite{glowinski1975approximation,gabay1976dual,boyd2011distributed}, 
suffers here from the same drawback as interior point methods --- at each iteration, a large linear system involving the matrices
$A_i$ must be solved, and this linear system is intractable.
Often the key to a successful application of ADMM is to exploit special problem structure
to solve this linear system efficiently, but it is not clear how to do that in this application.
A variant of ADMM known as linearized ADMM avoids the necessity of solving a linear system
at each iteration, and requires only matrix-vector multiplications using the matrices
$A_i$.  Related algorithms such as the Chambolle-Pock algorithm \cite{chambolle2011first,pock2009convex} have the same virtue.
However, in our investigation, the Chambolle-Pock algorithm was not able to solve
problem \eqref{generalFmoProb} to a sufficient level of accuracy in a reasonable amount
of time.  We encountered a similar difficulty when using ``block splitting'' versions
of ADMM \cite{parikh2014block} which replace a single large linear system with many small linear systems
which must be solved (in parallel) at each iteration.

Noting that there is a well known closed-form solution for the proximal operator
of the group sparsity penalty $\sum_b w_b \| x_b \|_2$, it may at first seem straightforward
to solve problem \eqref{generalFmoProb} using the proximal gradient method (also known
as the forward-backward method).  However, there is a challenge here as well, in that
we must be careful to handle the nonnegativity constraint on $x$ correctly.
The prior work on group sparsity for beam orientation optimization \cite{jia2011beam} used
the forward-backward method,
but the nonnegativity constraints on $x$ were enforced in a heuristic manner, without providing a theoretical justification.
In this section,
we show how to handle the nonnegativity constraints correctly, which allows us
to solve problem \eqref{generalFmoProb} efficiently using an accelerated
version of the proximal gradient method known as the Fast Iterative Shrinkage-Thresholding
Algorithm (FISTA) \cite{beck2009FISTA}.

\paragraph{Solution using FISTA}
FISTA solves convex optimization problems of the form
\begin{equation}
\label{fistaProb}
\mmz_{x \in \mathbb R^n} \, f(x) + g(x),
\end{equation}
where the convex function $f$ is assumed to be differentiable
(with a Lipschitz continuous gradient) and the convex function
$g$ is assumed to ``simple'' in the sense that its proximal operator
can be evaluated efficiently.  (We also require that $g$ is lower semi-continuous,
which is a mild assumption that is usually satisfied in practice.)
FISTA does not require $g$ to be differentiable.
Problem~\eqref{generalFmoProb} has the form \eqref{fistaProb},
where
\begin{align}
\label{fDef}
\notag f(x) &=  \sum_{i=0}^N \frac{\alpha_i}{2} \| (A_i x - d_i)_+\|_2^2 + \frac{\beta_i}{2} \| A_i x \|_2^2 \\
& \quad + \frac12\|(\ell - A_0 x)_+\|_2^2 + \gamma \| Dx \|_1^{(\mu)}
\end{align}
and 
\begin{equation}
\label{gDef}
g(x) = \begin{cases} \sum_{b=1}^B w_b \| x_b \|_2 & \quad \text{if } x \geq 0, \\
\infty & \quad \text{otherwise}.
\end{cases}
\end{equation}
The convex function $g$ enforces the constraint $x \geq 0$ by returning the value $\infty$
when this constraint is not satisfied.  (Enforcing hard constraints in this manner
is a standard technique in convex optimization.)

The key steps in each iteration of FISTA are to evaluate the gradient of
$f$ and the proximal operator of $g$.
To compute the gradient of $f$, we first note two facts that can be shown using basic calculus:
\begin{enumerate}
\item If $h(y) = \frac12 \| y_+ \|_2^2$, then $\nabla h(y) = y_+ = \max(y,0)$ (with maximum taken componentwise).
\item If $h$ is the Huber penalty function $h(y) = \| y \|_1^{(\mu)}$ (defined in equation~\eqref{huberDef}), then $\nabla h(y) = \frac{1}{\mu} P_{[-\mu,\mu]}(y)$,
where $P_{[-\mu,\mu]}(y)$ is the projection of the vector $y$ onto the set $\{ u \mid -\mu \leq u \leq \mu\}$.
(The inequalities are interpreted componentwise.)  Projecting onto this set is a simple componentwise ``clipping'' operation.
\end{enumerate}
It now follows from the chain rule that
\begin{align}
\label{gradfFormula}
\notag \nabla f(x) &=  \sum_{i=0}^N \alpha_i A_i^T(A_i x - d_i)_+ + \beta_i A_i^T A_i x \\
& \quad -A_0^T(\ell - A_0 x)_+ + \frac{\gamma}{\mu} D^T P_{[-\mu,\mu]}(Dx).
\end{align}

A formula for the prox-operator of $g$ 
is derived in appendix~\ref{proxAppendix}.  To state this formula, we first
express $g$ as $g(x) = \sum_{b=1}^B g_b(x_b)$,
where
\[
g_b(x_b) = \begin{cases} w_b \| x_b \|_2 &  \quad \text{if } x_b \geq 0, \\
\infty & \quad \text{otherwise}.
\end{cases}
\]
(Recall that $x_b$ is the fluence map for beam $b$, stored as a vector, and $x$ is the concatenation
of the vectors $x_b$.)
The prox-operator of $g$ is given by
\begin{align}
\label{keyProx}
\prox_{t g}(x )
&=  \begin{bmatrix} \prox_{t g_1}(x_1) \\ \prox_{t g_2}(x_2) \\ \vdots \\ \prox_{t g_B}(x_B) \end{bmatrix},\\
 \notag \prox_{t g_b}(x_b) &= \prox_{t w_b \| \cdot \|_2}(\max(x_b,0)).
\end{align}
Here $\prox_{t w_b \| \cdot \|_2}$ denotes the prox-operator of the $\ell_2$-norm
with parameter $t w_b$. A standard formula for the prox-operator of the $\ell_2$-norm states that
\[
\prox_{t w_b \| \cdot \|_2}(y) = y - Py,
\]
where $Py$ denotes the projection of $y$ onto $\ell_2$-norm ball of radius $t w_b$.
We have not found formula~\eqref{keyProx} for the prox-operator of $g$ elsewhere in the literature,
and the fact that there is a closed-form expression for the prox-operator
of $g$ in this case is a subtle but key point of this paper.
Without this formula we would be unable to use FISTA.
Using formulas~\eqref{gradfFormula} and~\eqref{keyProx} to compute the gradient of $f$ and the prox-operator
of $g$, it is now straightforward to solve problem \eqref{generalFmoProb} using FISTA
with line search (algorithm \ref{fistaAlg_lineSearch} in appendix \ref{fistaExplained}).

We next discuss two tricks to reduce the FISTA runtime.
\paragraph{Pruning beams}
In practice, FISTA (or any method with a similar convergence rate) tends to rule out most candidate beams very quickly.
An important trick to improve runtime is to occasionally throw out inactive
beams (and remove the corresponding columns from the matrices $A_i$).
This reduces the size of the optimization problem substantially.
In our implementation, we throw out inactive beams once every
$40$ iterations.  (We do not prune every iteration because there is a computational
expense associated with removing columns from a large sparse matrix, as a large
amount of data must be moved around in memory.  In our experiments, pruning every
$40$ iterations gave the greatest improvement in runtime.)
A beam is declared to be inactive if
its vector of beamlet intensities $x_b$
satisfies $\|x_b \|_2 < 10^{-6}$.
While this pruning step is strictly optional, we find that it
decreases runtime by a factor of approximately $4$ or $5$.
This is a standard
trick to improve runtime when solving optimization problems
with sparsity-inducing regularizers \cite{ghaoui2010safe,tibshirani2012strong,friedman2009glmnet}.

Although it is true that by pruning beams we are no longer guaranteed to find an optimal
solution to the optimization problem, in practice we find that beam pruning
has a negligible effect on which beams are selected.
For the four cases presented in section~\ref{numericalResults},
the same beams were selected both with and without pruning.
Methods to eliminate features while still guaranteeing a globally optimal solution
have been studied in the feature elimination literature~\cite{ghaoui2010safe,tibshirani2012strong},
and adapting these ``safe'' methods to beam orientation optimization is
a subject of future work.

\paragraph{Downsampling}
Due to the large number of candidate beams (typically $500 - 800$ in our experiments),
the matrices $A_i$ are huge and take up many gigabytes of computer memory.
To reduce memory requirements we sometimes uniformly downsample the voxel grid.
Specifically, in our experiments any structure larger than 10,000 voxels is downsampled uniformly
by a factor of 8 by keeping only the voxels $(i,j,k)$ where $i,j,$ and $k$ are multiples of $2$.
The corresponding rows of the matrices $A_i$ are omitted
and the vectors $\ell$ and $d_i$ are adjusted accordingly.
(In one case, referred to as ``H\&N'' below, we downsampled structures larger
than $10,000$ voxels by a factor of 12 by keeping only the voxels
$(i,j,k)$ where $i$ and $j$ are multiples of $2$ and $k$ is a multiple of $3$.)

\subsection{Experimental setup}
\label{setupSection}
\begin{table*}[htbp]
\renewcommand{\arraystretch}{1.3}
\caption{Prescription dose, PTV volume, number of candidate beams, and FISTA runtime for head and neck, lung, and prostate patients.}
\label{patientTable}
\centering
\begin{tabular}{l c c c c}
\toprule
Case & Prescription dose (Gy) & PTV volume (cc) & Number of candidate beams & FISTA runtime (min) \\ \midrule
H\&N   & 66 & 25.9 & 811 & 4.1\\ 
LNG \# 1 & 50 & 47.8 &  553 & 6.2\\ 
LNG \# 2 & 48 & 72.3 & 520 & 2.8\\ 
PRT  & 40 & 90.6 & 803 & 3.4\\ 
\bottomrule
\end{tabular}
\end{table*}

A head and neck case, two lung cases, and a prostate case were selected to test and evaluate the
proposed algorithm. 
The prescription doses and PTV volumes for each case are listed
in table~\ref{patientTable}.
In each case, we started with 1162 non-coplanar candidate beam firing positions distributed evenly
over the surface of a sphere, with roughly six degrees of separation between adjacent candidate beams. 
A 3D human surface measurement and a machine CAD model were utilized to map out the collision spaces, and
beam angles that resulted in collisions were removed. 
The details of collision space modeling were described previously \cite{victoria2015development}.
As a result, between 500 and 800 non-coplanar candidate beams were retained in each case for dose calculation and optimization
(see table~\ref{patientTable}).   
Beamlet dose was calculated for all beams within the conformal aperture~+5~mm margin 
using convolution/superposition with a 6~MV polyenergetic kernel \cite{neylon2014nonvoxel}. The dose calculation resolution
was isotropically~2.5~mm. The MLC leaf width at the isocenter was assumed to be 5~mm, identical to that of the clinical plans.

For each of the four cases, problem \eqref{generalFmoProb} was solved
using the FISTA with line search algorithm~\ref{fistaAlg_lineSearch} in appendix~\ref{fistaExplained}.
The parameters (weights) appearing in the penalty functions
and the vectors $\ell$ and $d_i$
were tuned (on a case by case basis) by trial and error to achieve high quality treatment plans.
The parameter $c$ in equation \eqref{weightFormula}
was chosen so that approximately $20$ beams
were active in the optimal solution to~\eqref{generalFmoProb}.
From these active beams, the $20$ with largest norm
$\| x_b \|_2$ were selected.  
Once these $20$ beams were selected, we performed a pure fluence map
optimization step, solving problem~\eqref{generalFmoProb} again
with the group sparsity term now omitted,
and using only the $20$ beams selected in the beam orientation
optimization step.
This pure FMO step is much faster than the beam angle selection
step because we only need to retain
the columns of $A$ corresponding to the $20$ selected beam angles;
in our Matlab implementation, this step usually takes about
$30$ seconds.  In both the beam orientation and fluence map optimization steps, 
a shell structure (of width $2$-$3$ cm) surrounding the PTV was included 
to penalize dose spillage to normal tissue.  
The resulting treatment plans were compared with clinical plans.

After the
first $50$ FISTA iterations, our FISTA implementation only attempted to increase the step size
on every fifth iteration.  In other words, after the $50$th iteration of FISTA,
the step
$t \coloneqq s \, t_{k-1}$ was only executed when $k$ was a multiple of $5$.
We took $r = s = 2$.  
Additionally, beams were pruned as discussed in section~\ref{booFistaSection}.
In each case FISTA was run for between 1000 and 2000 iterations, depending
on how many iterations were required for the number of active beams
to converge to a fixed value.
The optimization variable $x$ was initialized to all zeros.

For plan comparison, PTV D98, D99, and PTV homogeneity defined as D95/D5 were evaluated.
All treatment plans were scaled so that PTV D95 was equal to the prescription
dose.
OAR max and mean dose, denoted by $\text{D}_{\text{max}}^{\text{GS}}$ and $\text{D}_{\text{mean}}^{\text{GS}}$ for the group sparsity plan
 and $\text{D}_{\text{max}}^{\text{clinic}}$ and $\text{D}_{\text{mean}}^{\text{clinic}}$ for the clinical plan, 
were also calculated for assessment.  
For each OAR, the difference in max dose and the difference in mean dose between the two plans were computed. 
Max dose is defined as the dose at 2 percent of the structure volume, D2, which is recommended by the ICRU-83 report \cite{gregoire2011state}.


The group sparsity treatment plans were created on a computer
with two Intel~Xeon~CPU~E5-2687W~v3~3.10~GHz processors and 512 GB of RAM.
(This amount of RAM is not needed; the dose-calculation matrix typically
requires about 8 GB of RAM, after downsampling, in our experiments.)

\section{Results}
\label{numericalResults}
\begin{figure*}[htbp]
\centering
\includegraphics[width = \textwidth]{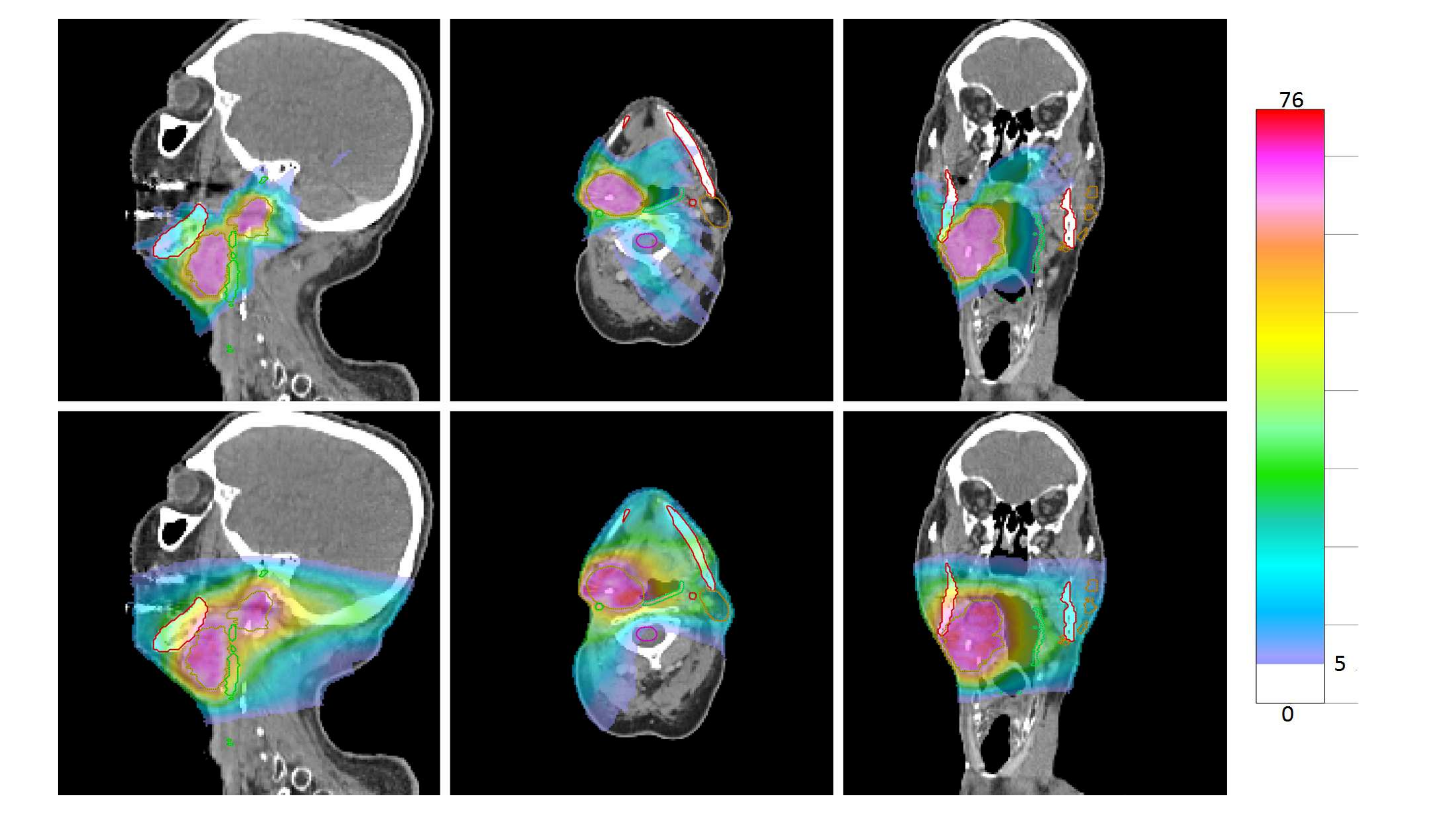}
\caption{Plans created for case ``H\&N''. The group sparsity plan is shown in the top row, and the clinical plan
is shown in the bottom row.  Dose below 5 Gy is not shown.}
\label{HNKLR_doseWash}
\end{figure*}

\begin{figure*}[htbp]
\centering
  \includegraphics[width=.8\linewidth]{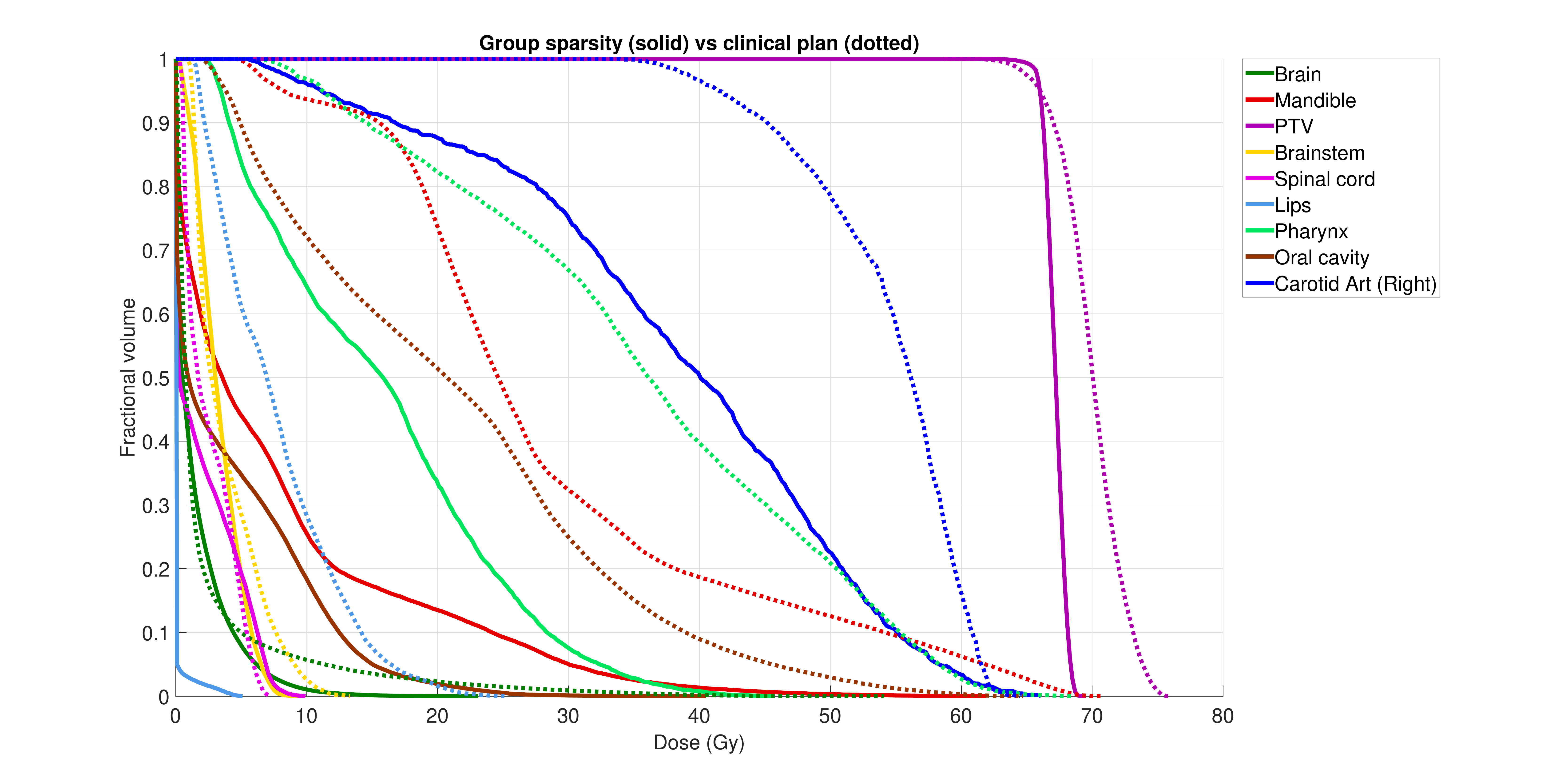}	
\caption{Dose-volume histogram for case ``H\&N''}
\label{dvh_HNKLR}
\end{figure*}

\begin{figure*}[htbp]
\centering
\includegraphics[width = \textwidth]{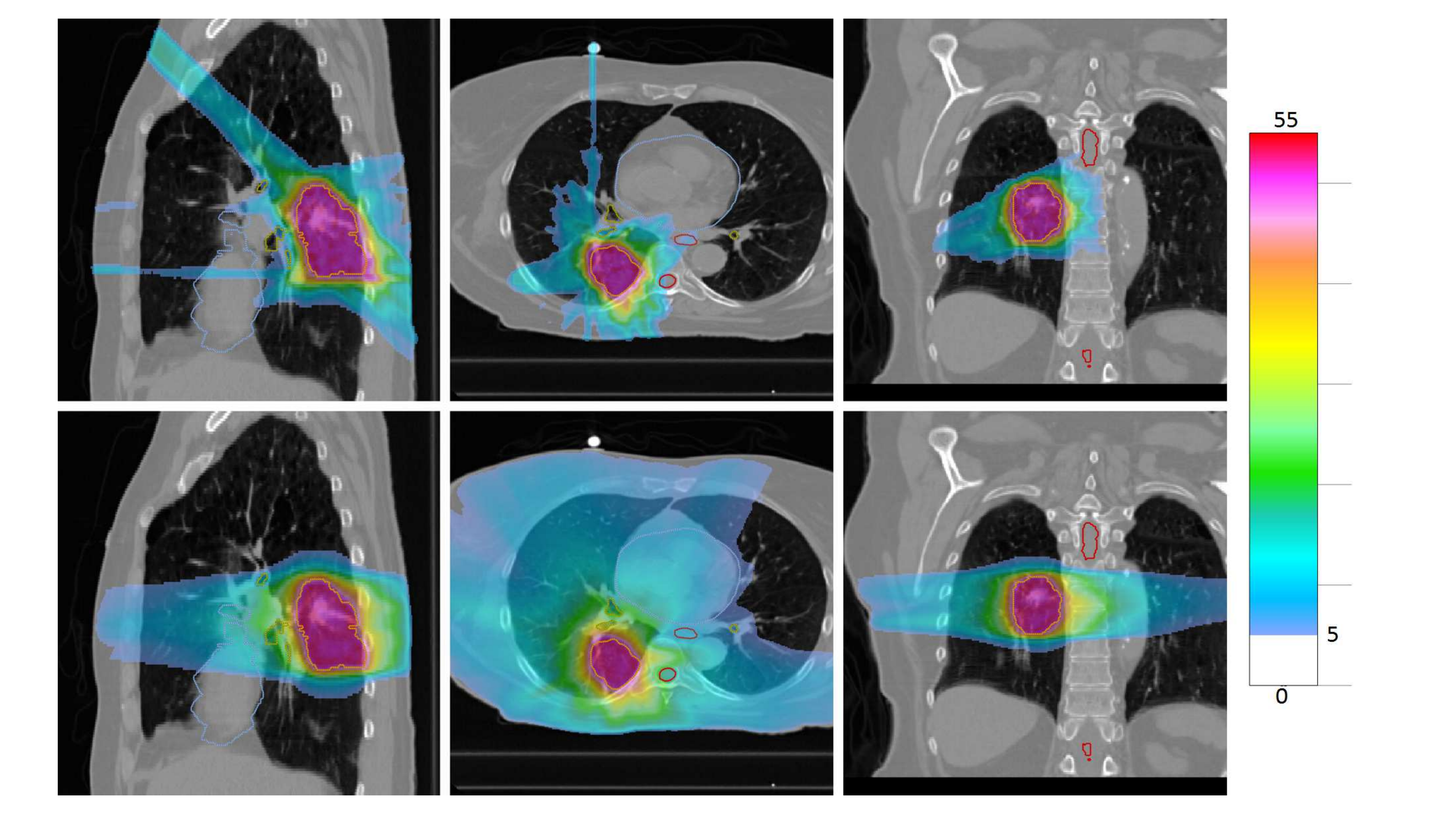}
\caption{Plans created for case ``LNG\#1''. The group sparsity plan is shown in the top row, and the clinical plan
is shown in the bottom row.  Dose below 5 Gy is not shown.}
\label{LNGND_doseWash}
\end{figure*}

\begin{figure*}[htbp]
\centering
  \includegraphics[width=.8\linewidth]{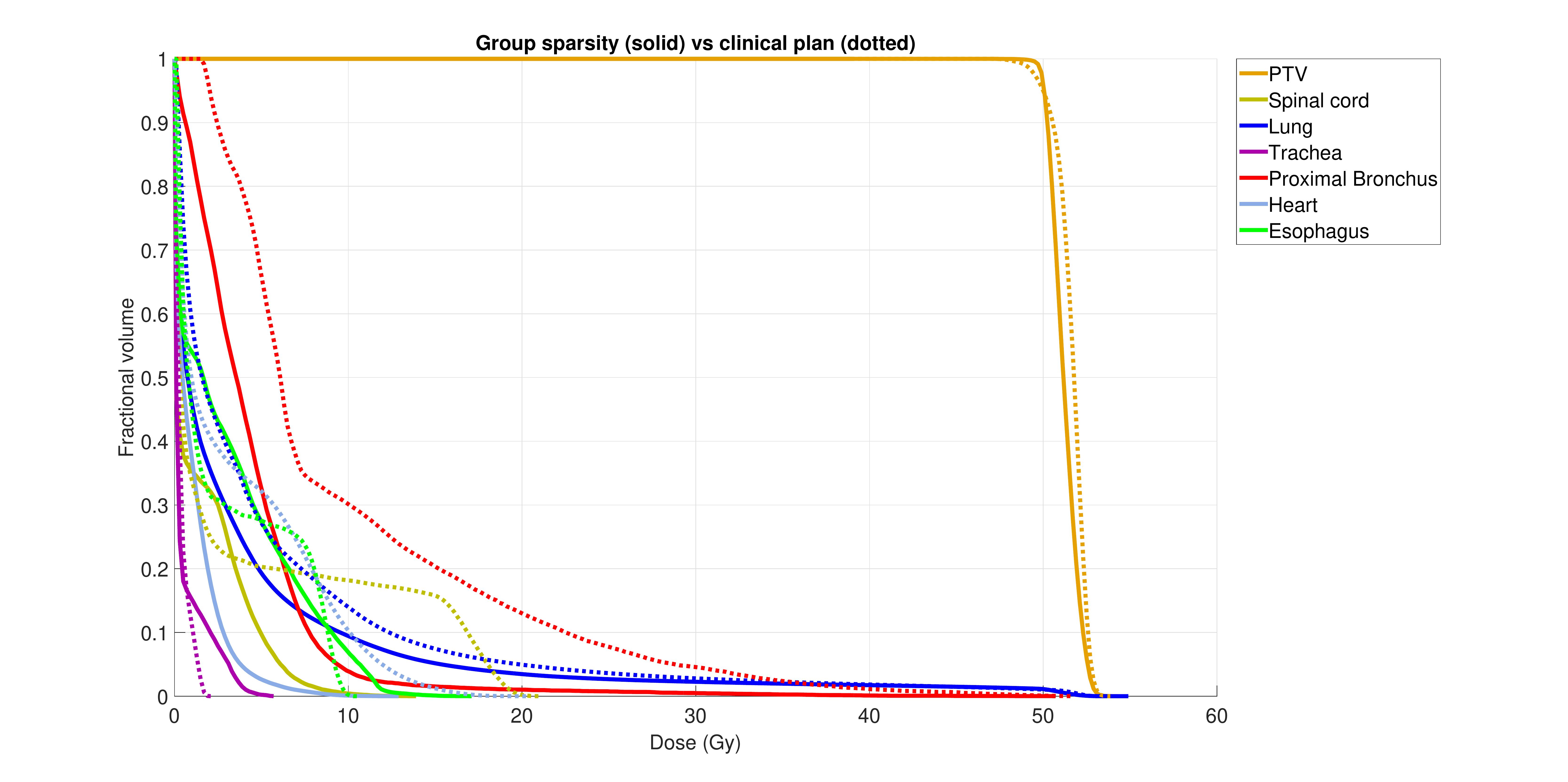}	
\caption{Dose-volume histogram for case ``LNG\#1''}
\label{dvh_LNGND}
\end{figure*}

Our goal in this section is to demonstrate that the group sparsity approach is now practical
for \emph{non-coplanar} IMRT --- the runtimes are reasonably fast and the dosimetric quality is
superior to that of the clinical plans we compare against.
Figures \ref{HNKLR_doseWash} and \ref{LNGND_doseWash} (top rows) show sagittal, transverse, and coronal views for non-coplanar treatment plans created
for the cases ``H\&N'' and ``LNG\#1'' using $20$
non-coplanar beams selected from $811$ and $553$ candidate beams, respectively, by our group sparsity approach.
Clinical plans for each case are shown in the bottom rows.
Corresponding dose-volume histograms for these cases are shown in figures \ref{dvh_HNKLR} and \ref{dvh_LNGND}.

\begin{table*}[htbp]
\renewcommand{\arraystretch}{1.3}
\caption{PTV coverage metrics for several cases.  Group sparsity results in black, 
clinical plan values in blue.  The homogeneity index HI is defined as $\text{D}95/\text{D}5$.}
\label{metrics}
\centering
\begin{tabular}{l l l l l l}
\toprule
Case & D95 (Gy) & D98 (Gy) & D99 (Gy) & $\text{D}_{\max}$& HI   \\ \midrule
H\&N  & 66.0 \color{blue} [66.0]& 65.7 \color{blue} [ 64.7] & 65.3 \color{blue} [63.9] & 68.6 \color{blue} [ 74.6] & .97 \color{blue} [.89]  \\
LNG \# 1 & 50.0 \color{blue} [50.0] & 49.9  \color{blue} [49.4] & 49.7  \color{blue} [49.0] & 52.8  \color{blue} [53.0] & .95 \color{blue} [.95]   \\
LNG \# 2 & 48.0 \color{blue} [48.0] & 47.7  \color{blue} [47.2] & 47.5 \color{blue} [46.7] & 51.3 \color{blue} [52.5] &.94 \color{blue} [.92]  \\
PRT & 40.0 \color{blue} [39.9] & 39.7 \color{blue} [39.6] &39.5 \color{blue} [39.3] &42.2 \color{blue} [41.9] &.95 \color{blue} [.96]  \\
\bottomrule
\end{tabular}
\vspace{2mm}
\caption{OAR dose differences for several cases. 
For each case, the difference in mean dose
$\text{D}_{\text{mean}}^{\text{GS}} - \text{D}_{\text{mean}}^{\text{clinic}}$
is computed for all OARs.
(``GS'' refers to the group sparsity treatment plan,
and ``clinic'' refers to the clinical treatment plan.)
The min, max, and average differences in mean dose are listed in columns 2 and 3.
Likewise, the min, max, and average values of
$\text{D}_{\text{max}}^{\text{GS}} - \text{D}_{\text{max}}^{\text{clinic}}$
are listed in columns 4 and 5.
}
\label{oarMetrics}
\begin{tabular}{l l l l l l l}
\toprule
&&\multicolumn{2}{c}{$\text{D}_{\text{mean}}^{\text{GS}} - \text{D}_{\text{mean}}^{\text{clinic}}$}& &\multicolumn{2}{c}{$\text{D}_{\max}^{\text{GS}} - \text{D}_{\max}^{\text{clinic}}$}\\
\midrule
Case &\quad & average (Gy) & range (Gy)& \qquad \qquad &  average (Gy) & range (Gy) \\ \midrule
H\&N && -10.4 & $[-21.1,-0.5]$& & -15.0 & $[-32.6,1.0]$ \\ 
LNG\#1  && -1.8 &$[-5.4,.30]$ && -7.1  &$[-23.6,2.2]$\\
LNG\#2  && -2.1 &$[-4.6,.13]$ && -5.2 &$[-13.7,1.1]$\\
PRT && -3.1 &$[-7.0,1.3]$ && -1.2 &$[-3.3,.02]$\\ 
\bottomrule
\end{tabular}
\end{table*}
 
Tables \ref{metrics} and \ref{oarMetrics} show treatment plan quality metrics for the four cases
listed in table~\ref{patientTable}.
The group sparsity plans show improvement in PTV D98 and PTV D99 for all cases,
with PTV D98 increasing on average by .53 Gy and
PTV D99 increasing on average by .78 Gy.
PTV homogeneity improved from .89 to .97 for case ``H\&N''
and remained constant or nearly constant for the other cases.
For case ``LNG\#1'', the R50 values were $3.20$ for the group sparsity plan
and $4.87$ for the clinical plan.
For case ``LNG\#2'', the R50 values were $2.80$ for the group sparsity plan and $7.51$ for the clinical plan.

The average and range of OAR dose differences for each case are reported in table~\ref{oarMetrics}. 
The group sparsity plans consistently show improvement over the clinical plans.
Considering all OARs for all cases, mean OAR dose was reduced by 7.7\% of the prescription dose, on average,
and max OAR dose was reduced by 11\% of the prescription dose, on average. 
Overall, considering doses washes, DVHs, and quality metrics,
the dosimetric quality of the plans created using the group sparsity approach is superior to
the dosimetric quality of the clinical plans.

\begin{figure*}[htbp]%
\centering
\subfigure[]{\includegraphics[width=.4\linewidth]{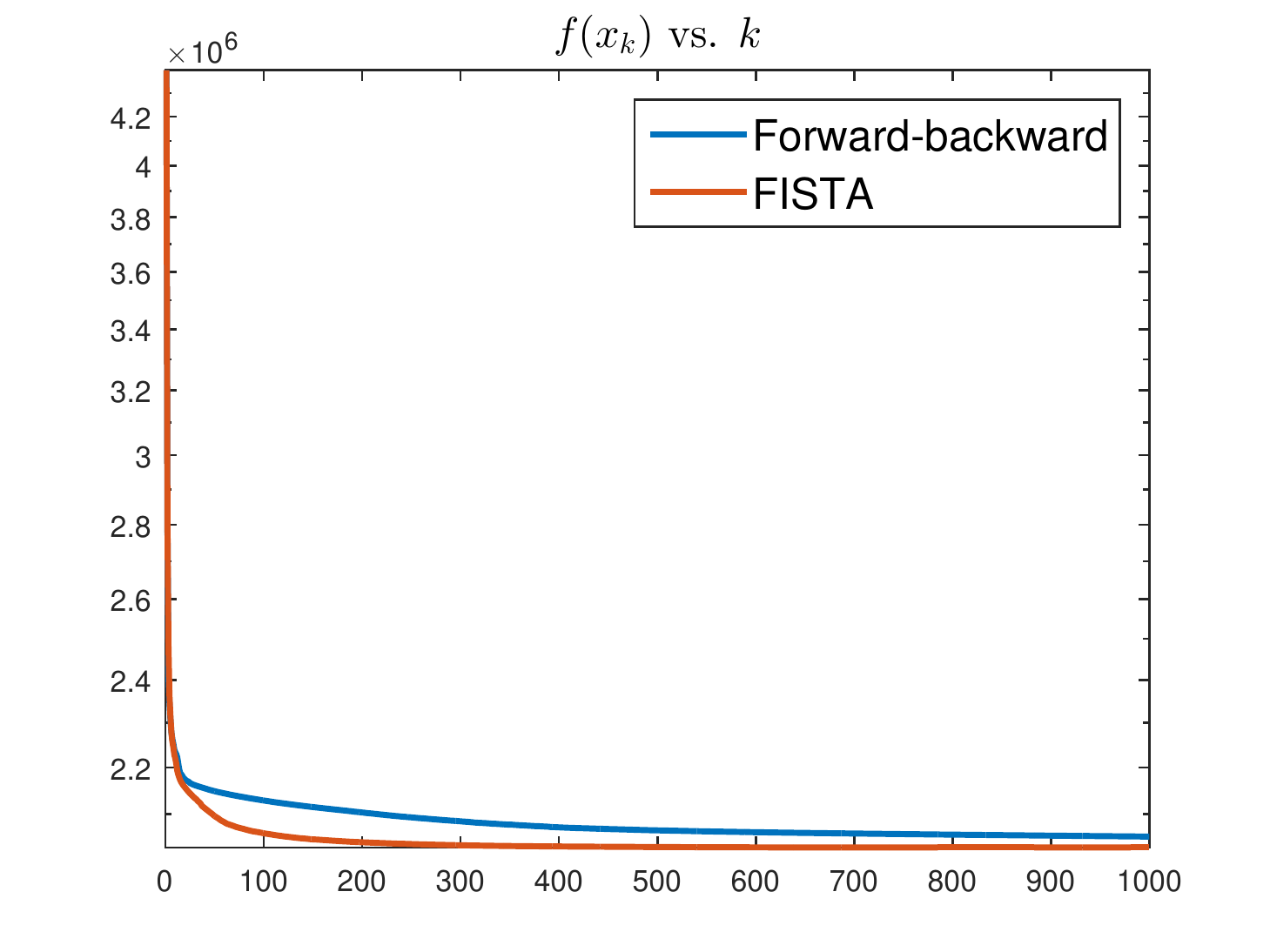}}	
\subfigure[]{\includegraphics[width=.4\linewidth]{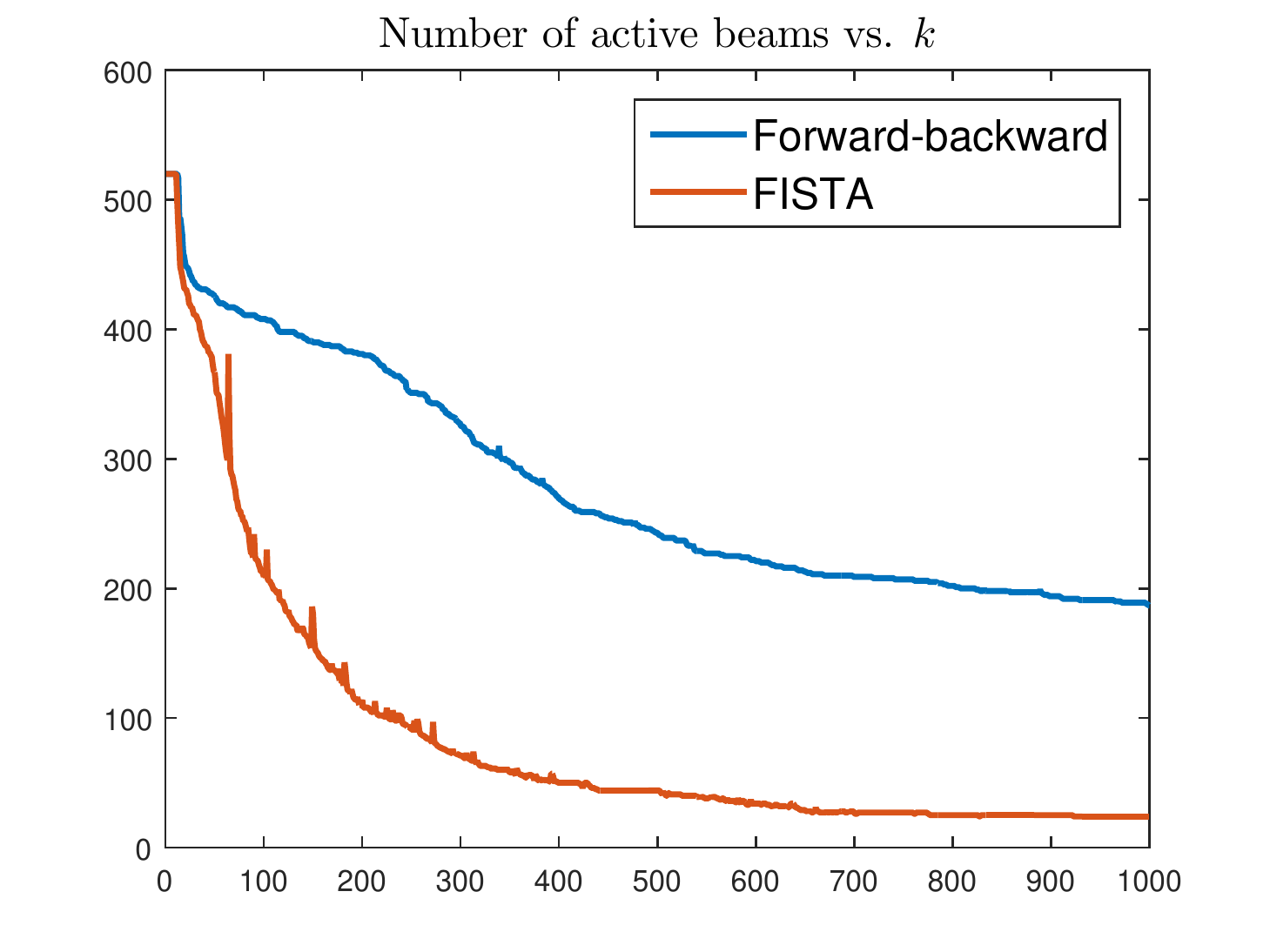}}
\caption{(a) Objective function value vs. iteration for both FISTA and the forward-backward method 
(which is the method used in \cite{jia2011beam}) for case ``LNG\#2''.
Notice that after only 200 iterations FISTA has already done much better than 1000 iterations
of the forward-backward method.
(b) Number of active beams at each iteration for FISTA and the forward-backward method.
After 1000 iterations, the solution computed by the forward-backward method still has $187$ active beams,
and the method has not converged.  Meanwhile, FISTA has converged to a solution with $24$ active beams (of which 
we kept the top $20$). The improved convergence rate of FISTA over the forward-backward method is the key to making
the group sparsity approach practical for non-coplanar cases.}
\label{convergencePlot}
\end{figure*} 

The FISTA runtimes for all cases are reported in table~\ref{patientTable}.
On average, the group sparsity plans took only 4.1 minutes to compute.
We have observed that the number of iterations required for FISTA to converge is not very sensitive
to the values of the weights appearing in the objective function.
The convergence plot for case ``LNG\#2'' in figure~\ref{convergencePlot} shows the necessity of 
using an \emph{accelerated} proximal gradient method. 
After 1000 iterations, the solution computed by the standard proximal gradient method
(also known as the forward-backward method, which was the method used in~\cite{jia2011beam}) still has $187$ active beams,
and the method has not converged.  Meanwhile, FISTA has converged to a solution with $24$ active beams (of which 
we kept the top $20$).  The improved convergence rate of FISTA over the forward-backward method
is the key to making the group sparsity approach practical for non-coplanar IMRT.

\section{Discussion}

The initial work \cite{jia2011beam} on group sparsity for beam orientation optimization
was not practical for non-coplanar IMRT due to the slow convergence of the optimization algorithm.
Without group sparsity, researchers developed greedy approaches in which
fluence map optimization steps were interleaved with beam angle selection steps (with
various heuristics available to select the next beam angle).
In this study, we adopted an accelerated proximal gradient method (specifically, FISTA) 
to make the group sparsity approach
practical for non-coplanar beam orientation optimization.
The $O(1/k^2)$ convergence rate of FISTA is a dramatic
improvement on the $O(1/k)$ convergence rate of the standard proximal gradient method,
also known as forward-backward method, which was used in \cite{jia2011beam}.
We have observed that in beam orientation optimization problems for $4\pi$
non-coplanar radiotherapy, the standard proximal gradient method (also known as forward-backward method) does
not converge to a group sparse solution in a reasonable amount of time;
thus it is essential to use an accelerated algorithm.
With the novel approach, we were able to solve large scale beam orientation optimization
problems in a few minutes.
Our Matlab implementation is able to select and optimize $20$ beams from $500-800$
candidate beams in about four minutes, producing treatment plans of dosimetric
quality superior to that of plans which were used in a clinical setting.
We improve considerably on the group sparsity results in \cite{jia2011beam},  which was restricted
to \emph{coplanar} beams and reported runtimes of a few hours for coplanar head and neck cases
with only 72 candidate beams.
There are potential benefits from the improved computational speed beyond making
beam orientation optimization practical for clinical adoption.
The improvement also enables integration of beam orientation optimization
with knowledge-based treatment planning and multi-criterion optimization
where a large number of plans need to be created, allowing
searching for optimal plans in an expanded solution space.

One of the challenges of promoting group sparsity is that
the $\ell_{2,1}$-norm is \emph{nondifferentiable}, so that we cannot use
classical optimization algorithms such as quasi-Newton methods which assume
the objective function is smooth.  Proximal algorithms 
such as FISTA are very well suited for this type of large scale,
nondifferentiable, constrained convex optimization problem.
A key contribution of this paper is that we provide a closed-form expression \eqref{keyProx}
for the prox-operator of the function $g$ given
by equation \eqref{gDef}.  (See appendix \ref{proxAppendix}
for our derivation of this expression, which we have not found
elsewhere in the literature.) Without the ability
to evaluate this prox-operator efficiently, 
the large scale group sparsity problem would be intractable.

The computational time can be further improved to near real time.
When solving problem \eqref{generalFmoProb} using FISTA, most of the
computational time is spent on matrix-vector multiplications
with the large sparse matrix $A$.  
Because these multiplications
are embarrassingly parallel, our algorithm should benefit greatly
from a multithreaded approach or a GPU implementation.  Preliminary experiments
suggest that these matrix-vector multiplications can be made over
$100$ times faster when performed on a GPU.
This is particularly important when a large number of plans need to be produced
for Pareto or adaptive radiotherapy planning.

\section{Conclusions}
\label{conclusions}

By using an accelerated proximal gradient method, enabled by the prox-operator
formula~\eqref{keyProx}, we have obtained an orders of magnitude
improvement on the runtimes reported in the initial work on beam orientation
optimization by group sparsity \cite{jia2011beam},
while improving on the dosimetric quality of plans that were used clinically.
We have demonstrated that the group sparsity approach is fast and effective
for \emph{non-coplanar} beam orientation optimization.

\appendix
\section{Proximal algorithms background}
\label{fistaSection}
 
In this section we define the proximal operator and briefly review
the proximal gradient method and FISTA.
An accessible introduction to proximal algorithms
can be found in \cite{parikh2013proximal}; see also \cite{chambolle2016introduction,236cNotes,combettes2011proximal}.

\subsection{Proximal operator.}
Let $f:\mathbb R^n \to \mathbb R \cup \{ \infty \}$
be a closed (i.e., lower semi-continuous) convex function.
The proximal operator (also known as ``prox-operator'') of $f$, with parameter $t > 0$, is defined by
\begin{equation}
\label{proxDef}
\prox_{tf}(x) = \argmin_u \quad f(u) + \frac{1}{2t} \|u - x \|_2^2.
\end{equation}
When we evaluate the proximal operator of $f$, it is as if we are trying to
reduce the value of $f$ without straying too far from $x$.
The parameter $t$ can be viewed as a ``step size'' that determines
how much we're penalized for moving away from $x$.
Proximal algorithms are iterative optimization algorithms that require
the evaluation of various prox-operators at each iteration.
For many important convex penalty functions, the prox-operator
has a simple closed-form expression and
can be evaluated very efficiently, at a computational
complexity that is linear in $n$.

\subsection{Proximal gradient method}

One of the most fundamental proximal algorithms, 
the proximal gradient method (also known as the \emph{forward-backward method})
solves optimization problems of the form
\begin{equation}
\label{proxGradProb}
\mmz \quad f(x) + g(x)
\end{equation}
where $f$ and $g$ are closed convex functions
and $f$ is differentiable with a Lipschitz continuous gradient.
The proximal gradient method with line search is recorded in algorithm \ref{proxGradAlg_lineSearch}.

\begin{algorithm*}
\caption{Proximal gradient method with line search} \label{proxGradAlg_lineSearch}
\begin{algorithmic}
\State Initialize $x_0$ and $t_0 > 0$, select $0 < r < 1, s > 1$
\For{$k = 1,2,\ldots$}
\State $t \coloneqq s \,t_{k-1}$
\MRepeat
\State $x \coloneqq \prox_{tg}(x_{k-1} - t \nabla f(x_{k-1}))$
\State \textbf{break if }$f(x) \leq f(x_{k-1}) + \langle \nabla f(x_{k-1}), x - x_{k-1} \rangle 
+ \frac{1}{2t} \| x - x_{k-1} \|_2^2$
\State $t \coloneqq r t$
\EndRepeat
\State $t_k \coloneqq t$
\State $x_k \coloneqq x$
\EndFor
\end{algorithmic}
\end{algorithm*}

\subsection{Accelerated proximal gradient methods}
\label{fistaExplained}
A recent theme in convex optimization research has been the development
of accelerated versions of the proximal gradient method
\cite{nesterov2004introductory,tseng2008accelerated,nesterov1983method,nesterov2005smooth,beck2009FISTA,beck2009gradient,beck2009fast,
scheinberg2014fast,nesterov2007gradient,becker2011nesta}.
These methods are popular for medical image reconstruction problems,
and they were applied to fluence map optimization problems 
(using the TFOCS software package \cite{becker2011templates}) in \cite{kim2012dose,kim2012efficient}.
In this paper, we focus on one particular accelerated method, 
FISTA \cite{beck2009FISTA} (short for ``fast iterative shrinkage-thresholding algorithm'').  
FISTA is an accelerated version of the proximal gradient method for solving problem \eqref{proxGradProb},
where (as before) $f$ and $g$ are closed convex functions, and
$f$ is differentiable with a Lipschitz continuous gradient.
The FISTA with line search algorithm is recorded in algorithm~\ref{fistaAlg_lineSearch}. 
Note that the FISTA iteration is only a minor modification of the proximal gradient
iteration.
Yet, FISTA converges at a rate of $O(1/k^2)$ (where $k$ is the iteration number),
whereas the proximal gradient iteration only converges at a rate
of $O(1/k)$.
FISTA's convergence rate of $O(1/k^2)$ is in some sense
optimal for a first-order method \cite{nesterov2004introductory}.
Although we focus on FISTA in this paper, 
it is not the only method that achieves this 
optimal $O(1/k^2)$ convergence rate \cite{tseng2008accelerated,nesterov2005smooth,nesterov2004introductory,becker2011nesta}.

\begin{algorithm*}
\caption{FISTA with line search} \label{fistaAlg_lineSearch}
\begin{algorithmic}
\State Initialize $x_0$ and $t_0 > 0$, set $v_0 \coloneqq x_0$, select $0 < r < 1$, $s > 1$
\For{$k = 1,2,\ldots$}
\State $t \coloneqq s \, t_{k-1}$
\MRepeat
\State $\theta \coloneqq \begin{cases} 1 & \quad \text{if } k = 1 \\
\text{positive root of } t_{k-1} \theta^2 = t \theta_{k-1}^2(1 - \theta) & \quad \text{if } k > 1 \end{cases}$
\State $y \coloneqq (1 - \theta) x_{k-1} + \theta v_{k-1}$
\State $x \coloneqq \prox_{t g}(y - t \nabla f(y))$
\State \textbf{break if } $f(x) \leq f(y) + \langle \nabla f(y),x - y \rangle + \frac{1}{2t} \| x - y \|_2^2$
\State $t \coloneqq r t$
\EndRepeat
\State $t_k \coloneqq t$
\State $\theta_k \coloneqq \theta$
\State $x_k \coloneqq x$
\State $v_k \coloneqq x_{k-1} + \frac{1}{\theta_k}(x_k - x_{k-1})$
\EndFor
\end{algorithmic}
\end{algorithm*}

\section{Prox-operator calculation}
\label{proxAppendix}
Here we derive a formula for the prox-operator of the function
\[
f(x) = \begin{cases} \|x\|_2 & \quad \text{if } x \geq 0 \\
\infty & \quad \text{otherwise}.
\end{cases}
\]
(The inequality $x \geq 0$ is interpreted componentwise.)  
Let $t > 0$. To evaluate $\prox_{tf}(\hat x)$,
we must find the minimizer for the problem
\begin{align}
\label{proxProb}
\mmz_x & \quad \|x\|_2 + \frac{1}{2t} \|x - \hat x \|_2^2 \\
\notag \subjto & \quad x \geq 0.
\end{align}
First note that if $\hat x_i \leq 0$ then there is no benefit
from taking $x_i$ to be positive.  If $x_i$ were positive,
then both terms in the objective function could be reduced just by setting
$x_i = 0$.

It remains only to select values for the other components of $x$.
This is a smaller optimization problem, with one unknown for each positive
component of $\hat x$.  The negative components of $\hat x$ are irrelevant
to the solution of this reduced problem.  Thus, we would still arrive at the same
final answer if the negative components of $\hat x$ were set equal to $0$
at the very beginning.

In other words, problem \eqref{proxProb} is equivalent to the problem
\begin{align*}
\mmz_x & \quad \|x\|_2 + \frac{1}{2t} \|x - \max(\hat x,0) \|_2^2 \\
\notag \subjto & \quad x \geq 0,
\end{align*}
which in turn is equivalent to the problem
\begin{align*}
\mmz_x & \quad \|x\|_2 + \frac{1}{2t} \|x - \max(\hat x,0) \|_2^2 \\
\end{align*}
(because there would be no benefit from taking any components of $x$
to be negative).  This shows that
\begin{equation}
\label{proxFormula}
\prox_{tf}(\hat x) = \prox_{t \| \cdot \|_2}(\max(\hat x,0)).
\end{equation}
As mentioned previously, a standard formula for the prox-operator of the $\ell_2$-norm
is 
\[
\prox_{t \| \cdot \|_2}(y) = y - Py,
\]
where $Py$ is the projection of $y$ onto the $\ell_2$-norm ball of radius $t$.


\section*{Acknowledgments}
This research was funded by NIH grants R43CA183390 and R01CA188300.
The initial work pertaining to fluence map optimization (but not group sparsity or beam orientation optimization)
was funded by RefleXion Medical.  The work pertaining to group sparsity and beam orientation optimization
was funded partially by Varian Medical Systems, Inc.
Thank you to Michael Grant who independently provided a derivation/proof of equation
\eqref{proxFormula}.







\end{document}